\documentclass[a4paper,12pt]{article}
\usepackage{latexsym}
\usepackage{amsmath}
\usepackage{amssymb}
\usepackage{amscd}

\newlength{\minitwocolumn}
\makeatletter
\@addtoreset{equation}{section}
\makeatother

\title{\bf 
The 19-Vertex Model\\
at critical regime $|q|=1$
}

\author{Takeo Kojima}
\date{\it
Department Mathematics,
College of Science and Technology,\\
Nihon University, Chiyoda-ku, Tokyo
101-0062, Japan\\~\\
{\rm \today}
}

\begin{document}
\maketitle

\begin{abstract}
We study the 19-vertex model
associated with the quantum group $U_q(\widehat{sl_2})$
at critical regime $|q|=1$.
We give the realizations of the type-I
vertex operators in terms of free bosons and free fermions.
Using these free fields realizations,
we give the integral representations for
the correlation functions.
\end{abstract}

~\\

\section{Introduction}

In this paper we shall study the 19 vertex model associated with
the quantum affine symmetry $U_q(\widehat{sl_2})$.
The 19-vertex model is a higher spin generalization of
the celebrated 6-vertex model,
whose
Boltzmann weights are given in
(\ref{Boltzmann}).
For the massive parameter case $|q|<1$,
M.Idzumi \cite{Id} derived the integral representations
of correlation functions from the viewpoint of
the representation thory of
the quantum group $U_q(\widehat{sl_2})$.
In this paper
we shall consider the problem at the critical regime $|q|=1$,
where the representation theory of
the quantum group $U_q(\widehat{sl_2})$
cannot be used.
M.Jimbo, H.Konno and T.Miwa \cite{JKM} studied
the 6-vertex model at critical regime $|q|=1$.
They presented
the free boson realizations of the vertex operators
and gave the trace constructions of the solutions
of the quantum Knizhnik-Zamolodchikov equations,
which represent the correlation functions.
In this paper we shall give the integral representations
for the correlation functions
of the 19-vertex model at critical regime $|q|=1$.
In order to give trace construction of the correlation
functions we need the free field realization of the
vertex operators. In this paper we present
the free field realizations of the vertex operators in
terms of free bosons and free fermions. 

The critical 19 vertex model is a limiting case of 
the fusion 8-vertex model.
The latter is massive, and the corner transfer method can
be applied \cite{JMN}.
Therefore we can conclude that
the correlation functions of the critical
19-vertex model are governed by the following
system of difference equations.
\begin{eqnarray}
&&G_{2N}(\cdots, \beta_{j+1}, \beta_j, \cdots)
_{\cdots \epsilon_{j+1} \epsilon_j \cdots}\nonumber\\
&=&\sum_{\epsilon_j' \epsilon_{j+1}'=0,1,2}
R_{\epsilon_j \epsilon_{j+1}}^{\epsilon_j' \epsilon_{j+1}'}
(\beta_{j}-\beta_{j+1})
G_{2N}(\cdots, \beta_{j}, \beta_{j+1}, \cdots)
_{\cdots \epsilon_{j}' \epsilon_{j+1}' \cdots},\\
&&G_{2N}(\beta_1 \cdots, \beta_{2N-1}, \beta_{2N}-i\lambda)
_{\epsilon_{1} \cdots \epsilon_{2N}}
=
G_{2N}(\beta_{2N}, \beta_1
\cdots, \beta_{2N-1})
_{\epsilon_{2N} \epsilon_1 \cdots \epsilon_{2N-1}}
,\nonumber
\\
\end{eqnarray}
where we the R-matrix is given in (\ref{Boltzmann}).
In this paper we set the deformation parameter $q$ as
following.
\begin{eqnarray}
q=\exp\left(-\frac{\pi i}{\xi}\right),~~~\xi>2.
\end{eqnarray}
We note that the above equations imply in particular the quantum
Knizhnik-Zamolodchikov equations.
In this paper we give the integral representations of the
above system of diffence equations.
The correlation functions are obtained by taking the shift
parameter $\lambda=2\pi$.

In this connection 
we should mention about the work \cite{L},
in which S.Lukyanov give the integral representations
of the form factors of sine-Gordon field theory.
We should mention about the work \cite{KY},
in which the authors gave the integral representations
of the correlation functions of the $U_q(\widehat{sl_n})$
analogue of the 6-vertex model at critical regime $|q|=1$. 
The special form factors 
of the $U_q(\widehat{sl_n})$
analogue of the 6-vertex model at critical regime $|q|=1$
are given by T.Miwa and Y.Takeyama \cite{MT}.

Now a few words about the organization of the paper.
In section 2 we formulate our problem.
In section 3 we give the free field realizations of the
vertex operators.
In section 4 we give proofs of properties of the free field
realizations.
In section 5 we give the integral representations of
the correlation functions.
In Appendix we summarize the multi-Gamma functions.

\section{Problem}
The purpose of this section is to formulate our problem.
At first we introduce two dimensional solvable lattice model,
so called 19-vertex model at critical regime $|q|=1$.
Consider an infinite square lattice consisting of oriented
lines, each carrying a spectral parameter varying fromline to 
line. The orientation of each line will be shown by
an arrow on it. A vertex is a crossing of two lines
with spectral parameters, say $\beta_1$ and $\beta_2$,
together with the adjacent 4 edges belonging to
the cross lines, as shown in Figure 1.
The edges are assigned spin-state variables :
$j_1,j_2,k_1,k_2$.
In the 19-vertex model,
each spin-state can take one of three different values $0,1,2$.
A spin configuration around the vertex is an assignment
of $0,1,2$ on the four edges.
There are 81 possible vertex configurations.
We assign each configuration a Boltzmann weight.
The set of all Boltzmann weight form the elements
of the $R$-matrix:
 
~\\

\unitlength 0.1in
\begin{picture}(20.25,23.30)(21.75,-28.45)
%
\special{pn 8}%
\special{pa 3400 800}%
\special{pa 3400 2400}%
\special{fp}%
\special{sh 1}%
\special{pa 3400 2400}%
\special{pa 3420 2333}%
\special{pa 3400 2347}%
\special{pa 3380 2333}%
\special{pa 3400 2400}%
\special{fp}%
\special{pa 4200 1600}%
\special{pa 2600 1600}%
\special{fp}%
\special{sh 1}%
\special{pa 2600 1600}%
\special{pa 2667 1620}%
\special{pa 2653 1600}%
\special{pa 2667 1580}%
\special{pa 2600 1600}%
\special{fp}%
\put(33.9000,-6.0000){\makebox(0,0){$k_1$}}%
\put(44.0000,-16.0000){\makebox(0,0){$k_2$}}%
\put(34.0000,-26.0000){\makebox(0,0){$j_1$}}%
\put(24.0000,-16.0000){\makebox(0,0){$j_2$}}%
\put(52.2000,-16.1000){\makebox(0,0){$=R_{j_1 j_2}^{k_1 k_2}(\beta_1-\beta_2)$}}%
\put(43.9000,-29.3000){\makebox(0,0){\bf Figure 1. Boltzmann weight}}%
\put(32.0000,-20.8000){\makebox(0,0){$\beta_1$}}%
\put(28.7000,-14.2000){\makebox(0,0){$\beta_2$}}%
\end{picture}%

\newpage
~\\
The matrix $R(\beta)$ acts on ${\mathbb{C}}^3 \otimes 
{\mathbb{C}}^3$ via the natural basis
$\{v_0,v_1,v_2\}$ of ${\mathbb{C}}^3$
as following.
\begin{eqnarray}
R(\beta)v_{k_1}\otimes v_{k_2}=
\sum_{j_1,j_2=0,1,2}
v_{j_1}\otimes v_{j_2}
R_{j_1 j_2}^{k_1 k_2}(\beta).
\end{eqnarray}
The Boltzmann weights are given explicitly by
\begin{eqnarray}
R(\beta)=\frac{1}{\kappa(\beta)}
\left(\begin{array}{ccccccccc}
1&&&&&&&&\\
&p&&e_2&&&&&\\
&&b&&g_2&&c_2&&\\
&e_1&&p&&&&&\\
&&h_1&&o&&h_2&&\\
&&&&&p&&e_2&\\
&&c_1&&g_1&&b&&\\
&&&&&e_1&&p&\\
&&&&&&&&1
\end{array}
\right).\label{Boltzmann}
\end{eqnarray}
Here we have set the normarized partition as
\begin{eqnarray}
\kappa(\beta)=-\frac{\displaystyle
{\rm sh}\left(\frac{1}{\xi}(\beta+\pi i)\right)}{
\displaystyle
{\rm sh}\left(\frac{1}{\xi}(\beta-\pi i)\right)
}.
\end{eqnarray}
The nonzero entries are given by
\begin{eqnarray}
p(\beta)&=&
-\frac{\displaystyle {\rm sh}\left(\frac{1}{\xi}\beta\right)}{
\displaystyle
{\rm sh}\left(\frac{1}{\xi}(\beta-2\pi i)\right)},\\
b(\beta)&=&
\frac{\displaystyle {\rm sh}\left(\frac{1}{\xi}\beta\right)
{\rm sh}\left(\frac{1}{\xi}(\beta+\pi i)\right)
}{\displaystyle
{\rm sh}\left(\frac{1}{\xi}(\beta-\pi i)\right)
{\rm sh}\left(\frac{1}{\xi}(\beta-2\pi i)\right)},
\\
o(\beta)&=&
\frac{\displaystyle
{\rm sh}\left(\frac{\pi i}{\xi}\right)
{\rm sh}\left(\frac{2 \pi i}{\xi}\right)
+
{\rm sh}\left(\frac{1}{\xi}\beta\right)
{\rm sh}\left(\frac{1}{\xi}(\beta-\pi i)\right)
}{
\displaystyle
{\rm sh}\left(\frac{1}{\xi}(\beta-\pi i)\right)
{\rm sh}\left(\frac{1}{\xi}(\beta-2\pi i)\right)
},
\end{eqnarray}
and
\begin{eqnarray}
c_1(\beta)&=&
\frac{\displaystyle \exp\left(\frac{2}{\xi}\beta\right)
{\rm sh}\left(\frac{\pi i}{\xi}\right)
{\rm sh}\left(\frac{2\pi i}{\xi}\right)}{
\displaystyle
{\rm sh}\left(\frac{1}{\xi}(\beta-\pi i)\right)
{\rm sh}\left(\frac{1}{\xi}(\beta-2 \pi i)\right)
},\\
e_1(\beta)&=&
-\frac{\displaystyle \exp\left(\frac{1}{\xi}\beta\right)
{\rm sh}\left(\frac{2\pi i}{\xi}\right)}{
\displaystyle
{\rm sh}\left(\frac{1}{\xi}(\beta-2\pi i)\right)},\\
c_2(\beta)&=&\exp\left(-\frac{4}{\xi}\beta \right)c_1(\beta),~~
e_2(\beta)=\exp\left(-\frac{2}{\xi}\beta\right),\\
g_1(\beta)&=&\frac{\displaystyle \exp\left(
\frac{1}{\xi}(\beta+\pi i)\right)
{\rm sh}\left(\frac{\pi i}{\xi} \right) 
{\rm sh}\left(\frac{1}{\xi}\beta\right)}
{\displaystyle 
{\rm sh}\left(\frac{1}{\xi}(\beta-\pi i)\right)
{\rm sh}\left(\frac{1}{\xi}(\beta-2 \pi i)\right)
},
\\
h_1(\beta)&=&
\frac{\displaystyle
2\exp\left(\frac{1}{\xi}(\beta-\pi i)\right)
{\rm sh}\left(\frac{1}{\xi}\beta\right)
{\rm sh}\left(\frac{2 \pi i}{\xi}\right)
{\rm ch}\left(\frac{\pi i}{\xi}\right)
}
{\displaystyle 
{\rm sh}\left(\frac{1}{\xi}(\beta-\pi i)\right)
{\rm sh}\left(\frac{1}{\xi}(\beta-2 \pi i)\right)
},\\
g_2(\beta)&=&\exp\left(-\frac{2}{\xi}\beta\right)g_1(\beta),~~~
h_2(\beta)=\exp\left(-\frac{2}{\xi}\beta\right)h_1(\beta).
\end{eqnarray}
~\\
The $R$-matrix satisfies the Yang-Baxter equation :
\begin{eqnarray}
R_{12}(\beta_1-\beta_2)R_{13}(\beta_1-\beta_3)
R_{23}(\beta_2-\beta_3)
=R_{23}(\beta_2-\beta_3)
R_{13}(\beta_1-\beta_3)R_{12}(\beta_1-\beta_2).
\end{eqnarray}
The 19-vertex model is a limiting case of 
the fusion 8-vertex model \cite{Takebe}.
The latter is massive, and the corner transfer matrix method 
can be applied \cite{JMN, Bax}.
Now we can conclude that the correlation functions
of the critical 19-vertex model are governed by
the following system of difference equations.
~\\
{\bf $R$-matrix Symmetry.}
\begin{eqnarray}
&&G_{2N}(\cdots, \beta_{j+1}, \beta_j, \cdots)
_{\cdots \epsilon_{j+1} \epsilon_j \cdots}\nonumber\\
&=&\sum_{\epsilon_j' \epsilon_{j+1}'=0,1,2}
R_{\epsilon_j \epsilon_{j+1}}^{\epsilon_j' \epsilon_{j+1}'}
(\beta_{j}-\beta_{j+1})
G_{2N}(\cdots, \beta_{j}, \beta_{j+1}, \cdots)
_{\cdots \epsilon_{j}' \epsilon_{j+1}' \cdots}.
\label{C1}
\end{eqnarray}
{\bf Cyclicity Condition.}
\begin{eqnarray}
G_{2N}(\beta_1 \cdots, \beta_{2N-1}, \beta_{2N}-i\lambda)
_{\epsilon_{1} \cdots \epsilon_{2N}}
=
G_{2N}(\beta_{2N}, \beta_1
\cdots, \beta_{2N-1})
_{\epsilon_{2N} \epsilon_1 \cdots \epsilon_{2N-1}}.
\label{C2}
\end{eqnarray}
The correlation functions :
\begin{eqnarray}
G_N(\beta_N'+\pi i, \cdots, \beta_1'+\pi i,
\beta_1, \cdots, \beta_N)_{2-j_N,\cdots,2-j_1,
j_1,\cdots, j_N}.
\end{eqnarray}
represents the configuration functions in Figure 2,
up to constant factors.

~\\

\unitlength 0.1in
\begin{picture}(38.00,30.10)(16.10,-31.15)
%
\special{pn 8}%
\special{pa 5410 620}%
\special{pa 1610 620}%
\special{fp}%
\special{sh 1}%
\special{pa 1610 620}%
\special{pa 1677 640}%
\special{pa 1663 620}%
\special{pa 1677 600}%
\special{pa 1610 620}%
\special{fp}%
\special{pa 5410 820}%
\special{pa 1610 820}%
\special{fp}%
\special{sh 1}%
\special{pa 1610 820}%
\special{pa 1677 840}%
\special{pa 1663 820}%
\special{pa 1677 800}%
\special{pa 1610 820}%
\special{fp}%
\special{pa 5410 2620}%
\special{pa 1610 2620}%
\special{fp}%
\special{sh 1}%
\special{pa 1610 2620}%
\special{pa 1677 2640}%
\special{pa 1663 2620}%
\special{pa 1677 2600}%
\special{pa 1610 2620}%
\special{fp}%
\special{pa 5410 2820}%
\special{pa 1610 2820}%
\special{fp}%
\special{sh 1}%
\special{pa 1610 2820}%
\special{pa 1677 2840}%
\special{pa 1663 2820}%
\special{pa 1677 2800}%
\special{pa 1610 2820}%
\special{fp}%
\special{pa 5010 420}%
\special{pa 5010 3020}%
\special{fp}%
\special{sh 1}%
\special{pa 5010 3020}%
\special{pa 5030 2953}%
\special{pa 5010 2967}%
\special{pa 4990 2953}%
\special{pa 5010 3020}%
\special{fp}%
\special{pa 4810 420}%
\special{pa 4810 3020}%
\special{fp}%
\special{sh 1}%
\special{pa 4810 3020}%
\special{pa 4830 2953}%
\special{pa 4810 2967}%
\special{pa 4790 2953}%
\special{pa 4810 3020}%
\special{fp}%
\special{pa 4610 420}%
\special{pa 4610 1420}%
\special{fp}%
\special{sh 1}%
\special{pa 4610 1420}%
\special{pa 4630 1353}%
\special{pa 4610 1367}%
\special{pa 4590 1353}%
\special{pa 4610 1420}%
\special{fp}%
\special{pa 4610 1820}%
\special{pa 4610 3020}%
\special{fp}%
\special{sh 1}%
\special{pa 4610 3020}%
\special{pa 4630 2953}%
\special{pa 4610 2967}%
\special{pa 4590 2953}%
\special{pa 4610 3020}%
\special{fp}%
\special{pa 4410 420}%
\special{pa 4410 1420}%
\special{fp}%
\special{sh 1}%
\special{pa 4410 1420}%
\special{pa 4430 1353}%
\special{pa 4410 1367}%
\special{pa 4390 1353}%
\special{pa 4410 1420}%
\special{fp}%
\special{pa 4410 1820}%
\special{pa 4410 3020}%
\special{fp}%
\special{sh 1}%
\special{pa 4410 3020}%
\special{pa 4430 2953}%
\special{pa 4410 2967}%
\special{pa 4390 2953}%
\special{pa 4410 3020}%
\special{fp}%
\special{pa 2010 420}%
\special{pa 2010 3020}%
\special{fp}%
\special{sh 1}%
\special{pa 2010 3020}%
\special{pa 2030 2953}%
\special{pa 2010 2967}%
\special{pa 1990 2953}%
\special{pa 2010 3020}%
\special{fp}%
\special{pa 2210 420}%
\special{pa 2210 3020}%
\special{fp}%
\special{sh 1}%
\special{pa 2210 3020}%
\special{pa 2230 2953}%
\special{pa 2210 2967}%
\special{pa 2190 2953}%
\special{pa 2210 3020}%
\special{fp}%
\special{pa 2410 420}%
\special{pa 2410 1420}%
\special{fp}%
\special{sh 1}%
\special{pa 2410 1420}%
\special{pa 2430 1353}%
\special{pa 2410 1367}%
\special{pa 2390 1353}%
\special{pa 2410 1420}%
\special{fp}%
\special{pa 2410 1820}%
\special{pa 2410 3020}%
\special{fp}%
\special{sh 1}%
\special{pa 2410 3020}%
\special{pa 2430 2953}%
\special{pa 2410 2967}%
\special{pa 2390 2953}%
\special{pa 2410 3020}%
\special{fp}%
\special{pa 2610 420}%
\special{pa 2610 1420}%
\special{fp}%
\special{sh 1}%
\special{pa 2610 1420}%
\special{pa 2630 1353}%
\special{pa 2610 1367}%
\special{pa 2590 1353}%
\special{pa 2610 1420}%
\special{fp}%
\special{pa 2610 1820}%
\special{pa 2610 3020}%
\special{fp}%
\special{sh 1}%
\special{pa 2610 3020}%
\special{pa 2630 2953}%
\special{pa 2610 2967}%
\special{pa 2590 2953}%
\special{pa 2610 3020}%
\special{fp}%
\put(24.1000,-16.2000){\makebox(0,0){$j_1$}}%
\put(46.1000,-16.2000){\makebox(0,0){$j_N$}}%
%
\special{pn 8}%
\special{pa 5410 620}%
\special{pa 1610 620}%
\special{fp}%
\special{sh 1}%
\special{pa 1610 620}%
\special{pa 1677 640}%
\special{pa 1663 620}%
\special{pa 1677 600}%
\special{pa 1610 620}%
\special{fp}%
\special{pa 5410 820}%
\special{pa 1610 820}%
\special{fp}%
\special{sh 1}%
\special{pa 1610 820}%
\special{pa 1677 840}%
\special{pa 1663 820}%
\special{pa 1677 800}%
\special{pa 1610 820}%
\special{fp}%
\special{pa 5410 2620}%
\special{pa 1610 2620}%
\special{fp}%
\special{sh 1}%
\special{pa 1610 2620}%
\special{pa 1677 2640}%
\special{pa 1663 2620}%
\special{pa 1677 2600}%
\special{pa 1610 2620}%
\special{fp}%
\special{pa 5410 2820}%
\special{pa 1610 2820}%
\special{fp}%
\special{sh 1}%
\special{pa 1610 2820}%
\special{pa 1677 2840}%
\special{pa 1663 2820}%
\special{pa 1677 2800}%
\special{pa 1610 2820}%
\special{fp}%
\special{pa 5010 420}%
\special{pa 5010 3020}%
\special{fp}%
\special{sh 1}%
\special{pa 5010 3020}%
\special{pa 5030 2953}%
\special{pa 5010 2967}%
\special{pa 4990 2953}%
\special{pa 5010 3020}%
\special{fp}%
\special{pa 4810 420}%
\special{pa 4810 3020}%
\special{fp}%
\special{sh 1}%
\special{pa 4810 3020}%
\special{pa 4830 2953}%
\special{pa 4810 2967}%
\special{pa 4790 2953}%
\special{pa 4810 3020}%
\special{fp}%
\special{pa 4610 420}%
\special{pa 4610 1420}%
\special{fp}%
\special{sh 1}%
\special{pa 4610 1420}%
\special{pa 4630 1353}%
\special{pa 4610 1367}%
\special{pa 4590 1353}%
\special{pa 4610 1420}%
\special{fp}%
\special{pa 4610 1820}%
\special{pa 4610 3020}%
\special{fp}%
\special{sh 1}%
\special{pa 4610 3020}%
\special{pa 4630 2953}%
\special{pa 4610 2967}%
\special{pa 4590 2953}%
\special{pa 4610 3020}%
\special{fp}%
\special{pa 4410 420}%
\special{pa 4410 1420}%
\special{fp}%
\special{sh 1}%
\special{pa 4410 1420}%
\special{pa 4430 1353}%
\special{pa 4410 1367}%
\special{pa 4390 1353}%
\special{pa 4410 1420}%
\special{fp}%
\special{pa 4410 1820}%
\special{pa 4410 3020}%
\special{fp}%
\special{sh 1}%
\special{pa 4410 3020}%
\special{pa 4430 2953}%
\special{pa 4410 2967}%
\special{pa 4390 2953}%
\special{pa 4410 3020}%
\special{fp}%
\special{pa 2010 420}%
\special{pa 2010 3020}%
\special{fp}%
\special{sh 1}%
\special{pa 2010 3020}%
\special{pa 2030 2953}%
\special{pa 2010 2967}%
\special{pa 1990 2953}%
\special{pa 2010 3020}%
\special{fp}%
\special{pa 2210 420}%
\special{pa 2210 3020}%
\special{fp}%
\special{sh 1}%
\special{pa 2210 3020}%
\special{pa 2230 2953}%
\special{pa 2210 2967}%
\special{pa 2190 2953}%
\special{pa 2210 3020}%
\special{fp}%
\special{pa 2410 420}%
\special{pa 2410 1420}%
\special{fp}%
\special{sh 1}%
\special{pa 2410 1420}%
\special{pa 2430 1353}%
\special{pa 2410 1367}%
\special{pa 2390 1353}%
\special{pa 2410 1420}%
\special{fp}%
\special{pa 2410 1820}%
\special{pa 2410 3020}%
\special{fp}%
\special{sh 1}%
\special{pa 2410 3020}%
\special{pa 2430 2953}%
\special{pa 2410 2967}%
\special{pa 2390 2953}%
\special{pa 2410 3020}%
\special{fp}%
\special{pa 2610 420}%
\special{pa 2610 1420}%
\special{fp}%
\special{sh 1}%
\special{pa 2610 1420}%
\special{pa 2630 1353}%
\special{pa 2610 1367}%
\special{pa 2590 1353}%
\special{pa 2610 1420}%
\special{fp}%
\special{pa 2610 1820}%
\special{pa 2610 3020}%
\special{fp}%
\special{sh 1}%
\special{pa 2610 3020}%
\special{pa 2630 2953}%
\special{pa 2610 2967}%
\special{pa 2590 2953}%
\special{pa 2610 3020}%
\special{fp}%
\put(24.1000,-16.2000){\makebox(0,0){$j_1$}}%
\put(46.1000,-16.2000){\makebox(0,0){$j_N$}}%
%
\special{pn 8}%
\special{pa 5410 620}%
\special{pa 1610 620}%
\special{fp}%
\special{sh 1}%
\special{pa 1610 620}%
\special{pa 1677 640}%
\special{pa 1663 620}%
\special{pa 1677 600}%
\special{pa 1610 620}%
\special{fp}%
\special{pa 5410 820}%
\special{pa 1610 820}%
\special{fp}%
\special{sh 1}%
\special{pa 1610 820}%
\special{pa 1677 840}%
\special{pa 1663 820}%
\special{pa 1677 800}%
\special{pa 1610 820}%
\special{fp}%
\special{pa 5410 2620}%
\special{pa 1610 2620}%
\special{fp}%
\special{sh 1}%
\special{pa 1610 2620}%
\special{pa 1677 2640}%
\special{pa 1663 2620}%
\special{pa 1677 2600}%
\special{pa 1610 2620}%
\special{fp}%
\special{pa 5410 2820}%
\special{pa 1610 2820}%
\special{fp}%
\special{sh 1}%
\special{pa 1610 2820}%
\special{pa 1677 2840}%
\special{pa 1663 2820}%
\special{pa 1677 2800}%
\special{pa 1610 2820}%
\special{fp}%
\special{pa 5010 420}%
\special{pa 5010 3020}%
\special{fp}%
\special{sh 1}%
\special{pa 5010 3020}%
\special{pa 5030 2953}%
\special{pa 5010 2967}%
\special{pa 4990 2953}%
\special{pa 5010 3020}%
\special{fp}%
\special{pa 4810 420}%
\special{pa 4810 3020}%
\special{fp}%
\special{sh 1}%
\special{pa 4810 3020}%
\special{pa 4830 2953}%
\special{pa 4810 2967}%
\special{pa 4790 2953}%
\special{pa 4810 3020}%
\special{fp}%
\special{pa 4610 420}%
\special{pa 4610 1420}%
\special{fp}%
\special{sh 1}%
\special{pa 4610 1420}%
\special{pa 4630 1353}%
\special{pa 4610 1367}%
\special{pa 4590 1353}%
\special{pa 4610 1420}%
\special{fp}%
\special{pa 4610 1820}%
\special{pa 4610 3020}%
\special{fp}%
\special{sh 1}%
\special{pa 4610 3020}%
\special{pa 4630 2953}%
\special{pa 4610 2967}%
\special{pa 4590 2953}%
\special{pa 4610 3020}%
\special{fp}%
\special{pa 4410 420}%
\special{pa 4410 1420}%
\special{fp}%
\special{sh 1}%
\special{pa 4410 1420}%
\special{pa 4430 1353}%
\special{pa 4410 1367}%
\special{pa 4390 1353}%
\special{pa 4410 1420}%
\special{fp}%
\special{pa 4410 1820}%
\special{pa 4410 3020}%
\special{fp}%
\special{sh 1}%
\special{pa 4410 3020}%
\special{pa 4430 2953}%
\special{pa 4410 2967}%
\special{pa 4390 2953}%
\special{pa 4410 3020}%
\special{fp}%
\special{pa 2010 420}%
\special{pa 2010 3020}%
\special{fp}%
\special{sh 1}%
\special{pa 2010 3020}%
\special{pa 2030 2953}%
\special{pa 2010 2967}%
\special{pa 1990 2953}%
\special{pa 2010 3020}%
\special{fp}%
\special{pa 2210 420}%
\special{pa 2210 3020}%
\special{fp}%
\special{sh 1}%
\special{pa 2210 3020}%
\special{pa 2230 2953}%
\special{pa 2210 2967}%
\special{pa 2190 2953}%
\special{pa 2210 3020}%
\special{fp}%
\special{pa 2410 420}%
\special{pa 2410 1420}%
\special{fp}%
\special{sh 1}%
\special{pa 2410 1420}%
\special{pa 2430 1353}%
\special{pa 2410 1367}%
\special{pa 2390 1353}%
\special{pa 2410 1420}%
\special{fp}%
\special{pa 2410 1820}%
\special{pa 2410 3020}%
\special{fp}%
\special{sh 1}%
\special{pa 2410 3020}%
\special{pa 2430 2953}%
\special{pa 2410 2967}%
\special{pa 2390 2953}%
\special{pa 2410 3020}%
\special{fp}%
\special{pa 2610 420}%
\special{pa 2610 1420}%
\special{fp}%
\special{sh 1}%
\special{pa 2610 1420}%
\special{pa 2630 1353}%
\special{pa 2610 1367}%
\special{pa 2590 1353}%
\special{pa 2610 1420}%
\special{fp}%
\special{pa 2610 1820}%
\special{pa 2610 3020}%
\special{fp}%
\special{sh 1}%
\special{pa 2610 3020}%
\special{pa 2630 2953}%
\special{pa 2610 2967}%
\special{pa 2590 2953}%
\special{pa 2610 3020}%
\special{fp}%
\put(24.1000,-16.2000){\makebox(0,0){$j_1$}}%
\put(46.1000,-16.2000){\makebox(0,0){$j_N$}}%
\put(61.8000,-17.6000){\makebox(0,0){\bf Figure 2. Correlators}}%
\put(24.1000,-32.0000){\makebox(0,0){$\beta_1'$}}%
\put(46.0000,-32.0000){\makebox(0,0){$\beta_N'$}}%
\put(46.0000,-2.0000){\makebox(0,0){$\beta_N$}}%
\put(24.1000,-1.9000){\makebox(0,0){$\beta_1$}}%
\end{picture}%

~\\

Now we can translate the problem to the following.\\
Find out the realizations of the vertex operators,
which satisfy the following conditions.\\
{\bf $R$-matrix Symmetry.}
\begin{eqnarray}
\Phi_{j_2}(\beta_2)\Phi_{j_1}(\beta_1)=
\sum_{k_1,k_2=0,1,2}R_{j_1 j_2}^{k_1 k_2}(\beta_1-\beta_2)
\Phi_{k_1}(\beta_1)\Phi_{k_2}(\beta_2).
\label{V1}
\end{eqnarray}
{\bf Homogeneity Condition.}
\begin{eqnarray}
e^{-\lambda D}\Phi_j(\beta)e^{\lambda D}=\Phi_j(\beta+i\lambda).
\label{V2}
\end{eqnarray}
Using the above vertex operators and the degree operator,
we can construct
the solutions of the system of difference equations
as following.
\begin{eqnarray}
&&G_{2N}(\beta_1,\cdots, \beta_{2N})_{j_1\cdots j_{2N}}=
\frac{{\rm tr}_{{\cal H}}\left(e^{-\lambda D}
\Phi_{j_1}(\beta_1) \cdots \Phi_{j_{2N}}(\beta_{2N})\right)}{
{\rm tr}_{{\cal H}}\left(e^{-\lambda D}\right)
}.
\end{eqnarray}
The $R$-matrix symmetry (\ref{C1}) follows from
the condition (\ref{V1}).
The cyclicity condition (\ref{C1}) follows from
the homogeneity condition (\ref{V2}).
In section 3 we give the free field realization of 
the vertex operators $\Phi_j(\beta)$.
In section 5 we give the degree operator $D$ and the 
space ${\cal H}$, on which the trace is evaluated.

\section{Free field realizations}
The purpose of this section is to give
the free field realization of the vertex operators.\\
Let us introduce the bose-fields by
\begin{eqnarray}
[b(t),b(t')]=-\frac{1}{t}
\frac{\displaystyle
{\rm sh}\left(\frac{\pi}{2}(\xi-2)t\right)}{
\displaystyle
{\rm sh}\left(\frac{\pi}{2}\xi t\right)}
\delta(t+t').
\end{eqnarray}
Let us set the basic operators by
\begin{eqnarray}
U_0(\beta)=:\exp\left(\int_{-\infty}^\infty b(t)e^{i\beta t}dt
\right):,\\
U_1(\beta)=:\exp\left(-\int_{-\infty}^\infty b(t)e^{i\beta t}dt
\right):
\end{eqnarray}
Let us set the fermion-fields by
\begin{eqnarray}
[\psi(t),\psi(t')]_+=2{\rm ch}(\pi t) \delta(t+t').
\end{eqnarray}
Fourier transformation of the fermion-field is given by 
\begin{eqnarray}
\widehat{\psi}(\beta)=\frac{1}{\sqrt{2\pi}}\int_{-\infty}^\infty
\psi(t)e^{it\beta}dt.
\end{eqnarray}
The free-field realizations of the vertex operators
are given by
\begin{eqnarray}
\Phi_2(\beta)&=&U_0(\beta),
\\
\Phi_1(\beta)&=&
\left(e^{\frac{\pi i}{\xi}}
+e^{-\frac{\pi i}{\xi}}\right)
\int_{-\infty}^\infty d\alpha~
\frac{\exp\left(\frac{1}{\xi}(\alpha-\beta)\right)}
{\displaystyle
{\rm sh}\left(\frac{1}{\xi}(\alpha-\beta+\pi i)\right)}\nonumber
\\
&&~~~~~~~~~~~~~~~~~~~~~~~~~~\times
U_0(\beta)U_1(\alpha)\widehat{\psi}(\alpha),
\\
\Phi_0(\beta)
&=&e^{\frac{\pi i}{\xi}}
\int_{-\infty}^\infty d\alpha_1 
\int_{-\infty}^\infty d\alpha_2
\prod_{k=1}^2
\frac{\exp\left(\frac{1}{\xi}(\alpha_k-\beta)\right)}
{\displaystyle
{\rm sh}\left(\frac{1}{\xi}
(\alpha_k-\beta+\pi i)\right)}\nonumber\\
&&~~~~~~~~~~~~~~~~~~~~\times
U_0(\beta)U_1(\alpha_1)U_1(\alpha_2)\widehat{\psi}(\alpha_1)
\widehat{\psi}(\alpha_2) 
\end{eqnarray}

\section{Proof}
The purpose of this section is to give proofs of
properties of vertex operators.
At first we explain the formulas of the form
\begin{eqnarray}
X(\beta_1)Y(\beta_2)=C_{XY}(\beta_1-\beta_2)
:X(\beta_1)X(\beta_2):,
\end{eqnarray}
where $X,Y=U_j$, and $C_{XY}(\beta)$ is a meromorphic
function on ${\mathbb{C}}$.
These formulae follow from the commutation relation
of the free bosons.
When we compute the contraction of the basic
operators,
we often encounter an integral
\begin{eqnarray}
\int_0^\infty
F(t)dt,
\end{eqnarray}
which is divergent at $t=0$.
Here we adopt the following prescription
for regularization :
it should be understood as the countour integral,
\begin{eqnarray}
\int_C F(t)\frac{{\rm log}(-t)}{2\pi i}dt, 
\end{eqnarray}
where the countour $C$ is given by
\\
~\\
~\\
\unitlength 0.1in
\begin{picture}(34.10,11.35)(17.90,-19.35)
%
\special{pn 8}%
\special{pa 5200 800}%
\special{pa 2190 800}%
\special{fp}%
\special{sh 1}%
\special{pa 2190 800}%
\special{pa 2257 820}%
\special{pa 2243 800}%
\special{pa 2257 780}%
\special{pa 2190 800}%
\special{fp}%
\special{pa 2190 1600}%
\special{pa 5190 1600}%
\special{fp}%
\special{sh 1}%
\special{pa 5190 1600}%
\special{pa 5123 1580}%
\special{pa 5137 1600}%
\special{pa 5123 1620}%
\special{pa 5190 1600}%
\special{fp}%
%
\special{pn 8}%
\special{pa 5190 1200}%
\special{pa 2590 1210}%
\special{fp}%
\put(25.9000,-12.1000){\makebox(0,0)[lb]{$0$}}%
%
\special{pn 8}%
\special{ar 2190 1210 400 400  1.5707963 4.7123890}%
\put(33.9000,-20.2000){\makebox(0,0){{\bf Contour} $C$}}%
\end{picture}%

~\\
The contractions of the basic operators have
the following forms.
\begin{eqnarray}
U_j(\beta_1)U_j(\beta_2)
=:U_j(\beta_1)U_j(\beta_2):
\frac{\displaystyle
\Gamma\left(i\frac{\beta_2-\beta_1}{\pi \xi}
+1-\frac{1}{\xi}\right)}{
\displaystyle
\Gamma\left(
i\frac{\beta_2-\beta_1}{\pi \xi}+\frac{1}{\xi}\right)}
\exp\left(\frac{\xi-2}{\xi}(\gamma+{\rm log}
(\pi \xi))\right).
\end{eqnarray}
Please see the Appendix.
The commutation relations of the basic operators are given by
\begin{eqnarray}
U_j(\beta_1)U_j(\beta_2)&=&
U_j(\beta_2)U_j(\beta_1)
\frac{
\displaystyle
{\rm sh}\left(\frac{1}{\xi}(\beta_1-\beta_2+\pi i)\right)}
{
\displaystyle
{\rm sh}\left(\frac{1}{\xi}(\beta_2-\beta_1+\pi i)\right)}
,~~(j=0,1),\\
U_0(\beta_1)U_1(\beta_2)&=&
U_1(\beta_2)U_0(\beta_1)
\frac{
\displaystyle
{\rm sh}\left(\frac{1}{\xi}(\beta_2-\beta_1+\pi i)\right)}
{
\displaystyle
{\rm sh}\left(\frac{1}{\xi}(\beta_1-\beta_2+\pi i)\right)}.
\end{eqnarray}
The anti-commutation relation becomes
\begin{eqnarray}
[\widehat{\psi}(\beta_1),\widehat{\psi}(\beta_2)]_+
=\delta(\beta_1-\beta_2+\pi i)+
\delta(\beta_2-\beta_1+\pi i).
\end{eqnarray}

Now let us start to prove the properties of
vertex operators.

~\\
{\it Proof of $R$-matrix symmetery}\\
Let us prove the equation :
\begin{eqnarray}
R_{21}^{12}(\beta_1-\beta_2)\Phi_1(\beta_1)\Phi_2(\beta_2)
+
R_{21}^{21}(\beta_1-\beta_2)\Phi_2(\beta_1)\Phi_1(\beta_2)
=\Phi_{1}(\beta_2)\Phi_2(\beta_1).\label{proof1}
\end{eqnarray}
By using the commutation relations of basic operators,
we can rearrange the operator part as
\begin{eqnarray}
U_0(\beta_1)U_0(\beta_2)U_1(\alpha)\widehat{\psi}(\alpha).
\end{eqnarray}
The equation (\ref{proof1}) follows from
the integrand identity :
\begin{eqnarray}
-e_1(\beta_1-\beta_2)e^{-\frac{1}{\xi}\beta_1}{\rm sh}\left(
\frac{1}{\xi}(\beta_2-\alpha+\pi i)
\right)+p(\beta_1-\beta_2)e^{-\frac{1}{\xi}\beta_2}
{\rm sh}\left(
\frac{1}{\xi}(\beta_1-\alpha-\pi i)\right)
\nonumber\\
=e^{-\frac{1}{\xi}\beta_2}
{\rm sh}\left(
\frac{1}{\xi}(\beta_1-\alpha+\pi i)\right).
\end{eqnarray}
Let us prove the equation :
\begin{eqnarray}
R_{11}^{02}(\beta_1-\beta_2)\Phi_0(\beta_1)\Phi_2(\beta_2)+
R_{11}^{11}(\beta_1-\beta_2)\Phi_1(\beta_1)\Phi_1(\beta_2)
\nonumber
\\
+
R_{11}^{20}(\beta_1-\beta_2)\Phi_2(\beta_1)\Phi_0(\beta_2)
=
\Phi_1(\beta_2)
\Phi_1(\beta_1).\label{proof2}
\end{eqnarray}
By using the commutation relations of basic operators,
we can rearrange the operator part as
\begin{eqnarray}
U_0(\beta_1)U_0(\beta_2)U_1(\alpha_1)U_1(\alpha_2)
\widehat{\psi}(\alpha_1)\widehat{\psi}(\alpha_2).
\end{eqnarray}
Let us set
\begin{eqnarray}
H(\alpha)=\frac{\displaystyle
{\rm sh}\left(\frac{1}{\xi}(\alpha+\pi i)\right)}
{\displaystyle
{\rm sh}\left(\frac{1}{\xi}(-\alpha+\pi i)
\right)}.
\end{eqnarray}
Consider the integral of the form :
\begin{eqnarray}
\int_{-\infty}^\infty
d\alpha_1
\int_{-\infty}^\infty
d\alpha_2
F(\alpha_1,\alpha_2)U_1(\alpha_1)U_1(\alpha_2)
\widehat{\psi}(\alpha_1)
\widehat{\psi}(\alpha_2).
\end{eqnarray}
Due to the commutation relation of $U_1(\alpha)$ and the
anti-commutation relation of $\widehat{\psi}(\alpha)$,
the above integral equals to
\begin{eqnarray}-
\int_{-\infty}^\infty
d\alpha_1
\int_{-\infty}^\infty
d\alpha_2 H(\alpha_2-\alpha_1)
F(\alpha_1,\alpha_2)U_1(\alpha_1)U_1(\alpha_2)
\widehat{\psi}(\alpha_1)
\widehat{\psi}(\alpha_2)\nonumber
\\
-H(\pi i)\int_{-\infty}^\infty d\alpha
F(\alpha-\pi i,\alpha)U_1(\alpha)U_1(\alpha-\pi i)
,
\end{eqnarray}
where we have used the relation:
$H(-\pi i)=0$. Note that the part
$H(\pi i)U_1(\alpha)U_1(\alpha-\pi i)$
is convergent.\\
Observing this we define 'weak equality' in the following sense.
We say that the function $G_1(\alpha_1,\alpha_2)$
and $G_2(\alpha_1,\alpha_2)$
are equal in weak sense if
\begin{eqnarray}
G_1(\alpha_1,\alpha_2)-H(\alpha_2-\alpha_1)
G_1(\alpha_2,\alpha_1)=
G_2(\alpha_1,\alpha_2)-H(\alpha_2-\alpha_1)
G_2(\alpha_2,\alpha_1).
\end{eqnarray}
We write 
\begin{eqnarray}
G_1(\alpha_1,\alpha_2)\sim
G_2(\alpha_1,\alpha_2).
\end{eqnarray}
Note that the equation
\begin{eqnarray}
G_1(\alpha-\pi i,\alpha)=
G_2(\alpha-\pi i,\alpha),
\end{eqnarray}
is a special case of weakly equality.
In oreder to prove the equation (\ref{proof2})
it is enough to prove the equality of the integrand part
in weakly sense.
The equation (\ref{proof2}) follows from the following weakly sense identity.
\begin{eqnarray}
h_1(\beta_1-\beta_2)
e^{-\frac{\pi i}{\xi}}
{\rm sh}
\left(\frac{1}{\xi}(\beta_2-\alpha_1+\pi i)
\right)
{\rm sh}\left(
\frac{1}{\xi}(\beta_2-\alpha_2+\pi i)\right)
\nonumber\\
+o(\beta_1-\beta_2)
(1+e^{-\frac{2\pi i}{\xi}})^2
{\rm sh}
\left(\frac{1}{\xi}(\beta_2-\alpha_1+\pi i)
\right)
{\rm sh}\left(
\frac{1}{\xi}(\alpha_2-\beta_1+\pi i)\right)
\nonumber\\
+h_2(\beta_1-\beta_2)
e^{-\frac{\pi i}{\xi}}
{\rm sh}\left(
\frac{1}{\xi}(\alpha_1-\beta_1+\pi i)\right)
{\rm sh}
\left(\frac{1}{\xi}(\alpha_2-\beta_1+\pi i)
\right)\nonumber\\
\sim
(1+e^{-\frac{2\pi i}{\xi}})^2
{\rm sh}\left(
\frac{1}{\xi}(\beta_1-\alpha_1+\pi i)\right)
{\rm sh}
\left(\frac{1}{\xi}(\alpha_2-\beta_2+\pi i)
\right).\label{weak}
\end{eqnarray}
As the same arguments as the above we obtain the
commutation relation of the vertex operators (\ref{V1}).

\section{Correlation functions}
In this section we derive a solution of the system of difference
equations (\ref{C1}) and (\ref{C2}), algebraically,
and obtain an integral representation of it.\\
Let us introduce the Fock space ${\cal H}^b$ generated by
$|vac \rangle_b$ which satisfies
\begin{eqnarray}
b(t)|vac\rangle_b=0,~~{\rm if}~~t>0.
\end{eqnarray}
Let us introduce the Fock space ${\cal H}^\psi$ generated by
$|vac \rangle_\psi$ which satisfies
\begin{eqnarray}
\psi(t)|vac\rangle_\psi=0,~~{\rm if}~~t>0.
\end{eqnarray}
Set the space ${\cal H}$ by
\begin{eqnarray}
{\cal H}={\cal H}^b \otimes {\cal H}^\psi.
\end{eqnarray}
Let us introduce the degree operators $D^b$ and $D^\psi$ by
\begin{eqnarray}
D^b b(-t) |vac \rangle_b= t b(-t)|vac \rangle_b,~~
D^\psi \psi(-t) |vac \rangle_\psi
= t \psi(-t)|vac \rangle_\psi,~~t>0.
\end{eqnarray}
Set the degree operator $D$ on ${\cal H}$ by
\begin{eqnarray}
D=D^b \otimes id+id \otimes D^\psi.
\end{eqnarray}
We have
\begin{eqnarray}
e^{\lambda D}U_j(\beta)e^{-\lambda D}=U_j(\beta+i\lambda),
~~
e^{\lambda D}\widehat{\psi}(\beta)e^{-\lambda D}=\widehat{\psi}
(\beta+i\lambda).
\end{eqnarray}
Therefore the vertex operators satisfy
the homogeneity condition.
\begin{eqnarray}
e^{-\lambda D}\Phi_j(\beta)e^{\lambda D}=\Phi_j(\beta+i\lambda).
\end{eqnarray}
Now let us consider the trace function for $\lambda>0$ 
defined by
\begin{eqnarray}
&&G_{2N}(\beta_1,\cdots, \beta_{2N})_{j_1\cdots j_{2N}}=
\frac{{\rm tr}_{{\cal H}}\left(e^{-\lambda D}
\Phi_{j_1}(\beta_1) \cdots \Phi_{j_{2N}}(\beta_{2N})\right)}{
{\rm tr}_{{\cal H}}\left(e^{-\lambda D}\right)
}.
\end{eqnarray}
The trace of the bosonic parts is evaluated as followings.
\begin{eqnarray}
\frac{{\rm tr}_{{\cal H}^b}\left(e^{-\lambda D^b}
b(t)b(t')\right)}{
{\rm tr}_{{\cal H}^b}\left(e^{-\lambda D^b}\right)
}=\frac{e^{\lambda t}}{e^{\lambda t}-1}
[b(t),b(t')].
\end{eqnarray}
The product of the basic operator $U_j(\beta),~(j=0,1)$ is
evaluated as
\begin{eqnarray}
&&\frac{{\rm tr}_{{\cal H}^b}\left(e^{-\lambda D^b}
U_j(\beta_1)U_j(\beta_2)\right)}{
{\rm tr}_{{\cal H}^b}\left(e^{-\lambda D^b}\right)
}= Const.
~\varphi_1(\beta_1-\beta_2)\nonumber\\
&\times&
{\rm sh}\left(\frac{\pi}{\lambda}(\beta_1-\beta_2+\pi i)\right)
{\rm sh}\left(\frac{\pi}{\lambda}(\beta_1-\beta_2-\pi i)\right)
{\rm sh}\left(\frac{1}{\xi}(\beta_1-\beta_2+\pi i)\right),
\end{eqnarray}
where we set the kernel function as
\begin{eqnarray}
\varphi_1(\beta)=\frac{1}{
S_2(i\beta-\pi|\pi \xi, \lambda)
S_2(-i\beta-\pi|\pi \xi, \lambda)
}.
\end{eqnarray}
The product of the basic operators $U_0(\beta)$ and $U_1(\alpha)$
is evaluated as
\begin{eqnarray}
&&\frac{{\rm tr}_{{\cal H}^b}\left(e^{-\lambda D^b}
U_0(\beta)U_1(\alpha)\right)}{
{\rm tr}_{{\cal H}^b}\left(e^{-\lambda D^b}\right)
}= Const.
~\varphi_2(\beta-\alpha)~
{\rm sh}\left(\frac{1}{\xi}(\beta-\alpha-\pi i)\right),
\end{eqnarray}
where we set the kernel function as
\begin{eqnarray}
\varphi_2(\alpha)=\frac{1}{
S_2(i\alpha+\pi|\pi \xi, \lambda)
S_2(-i\alpha+\pi|\pi \xi, \lambda)
}.
\end{eqnarray}
The trace of the fermionic parts is evaluated as followings.
\begin{eqnarray}
\frac{{\rm tr}_{{\cal H}^\psi}
\left(e^{-\lambda D^\psi}
\psi(t)\psi(t')\right)}{
{\rm tr}_{{\cal H}^\psi}
\left(e^{-\lambda D^\psi}\right)
}=\frac{e^{\lambda t}}{e^{\lambda t}+1}
[\psi(t),\psi(t')]_+.
\end{eqnarray}
Let us set the auxiliary function 
${\cal J}(\alpha)$ by
\begin{eqnarray}
{\cal J}(\alpha)=\frac{1}{2\pi}
\int_{-\infty}^\infty
\frac{e^{i\alpha t}}{1+e^{-\lambda t}} dt.
\end{eqnarray}
We then have
\begin{eqnarray}
\frac{{\rm tr}_{{\cal H}^\psi}
\left(e^{-\lambda D^\psi}
\widehat{\psi}(\alpha_1)
\widehat{\psi}(\alpha_2)\right)}{
{\rm tr}_{{\cal H}^\psi}
\left(e^{-\lambda D^\psi}\right)
}={\cal J}(\alpha_1-\alpha_2+\pi i)+
{\cal J}(\alpha_1-\alpha_2-\pi i).
\end{eqnarray}

The trace of the vertex operators is evaluated by
applying the Wick's theorem.\\
The one-point correlation functions are evaluated as follows.

~\\

\begin{eqnarray}
G_{2}(\beta_1,\beta_2)_{2,0}&=&
e^{-\frac{2}{\xi}\beta_2}
\frac{
S_2(i(\beta_2-\beta_1)+\pi|\pi \xi, \lambda)}{
S_2(i(\beta_2-\beta_1)+\pi \xi-\pi|\pi \xi, \lambda)}
\int_{-\infty}^\infty d\alpha_1
\int_{-\infty}^\infty d\alpha_2 \nonumber\\
&\times&\left\{\prod_{j,k=1}^2 \varphi_2(\beta_j-\alpha_k)\right\}
\varphi_1(\alpha_1-\alpha_2)({\cal J}(\alpha_1-\alpha_2+\pi i)
+{\cal J}(\alpha_1-\alpha_2-\pi i))
\nonumber\\
&\times&
e^{\frac{1}{\xi}(\alpha_1+\alpha_2)}
{\rm sh}\left(\frac{\pi}{\lambda}(\alpha_1-\alpha_2+\pi i)\right)
{\rm sh}\left(\frac{\pi}{\lambda}(\alpha_2-\alpha_1+\pi i)\right)
\nonumber\\
&\times&
{\rm sh}\left(\frac{1}{\xi}(\alpha_1-\alpha_2+\pi i)\right)
{\rm sh}\left(\frac{1}{\xi}(\alpha_1-\beta_1+\pi i)\right)
{\rm sh}\left(\frac{1}{\xi}(\alpha_2-\beta_1+\pi i)\right)
,\nonumber\\
\end{eqnarray}

~\\

\begin{eqnarray}
G_{2}(\beta_1,\beta_2)_{2,0}&=&
e^{-\frac{2}{\xi}\beta_1}
\frac{
S_2(i(\beta_2-\beta_1)+\pi|\pi \xi, \lambda)}{
S_2(i(\beta_2-\beta_1)+\pi \xi-\pi|\pi \xi, \lambda)}
\int_{-\infty}^\infty d\alpha_1
\int_{-\infty}^\infty d\alpha_2 \nonumber\\
&\times&\left\{\prod_{j,k=1}^2 \varphi_2(\beta_j-\alpha_k)\right\}
\varphi_1(\alpha_1-\alpha_2)({\cal J}(\alpha_1-\alpha_2+\pi i)
+{\cal J}(\alpha_1-\alpha_2-\pi i))
\nonumber\\
&\times&
e^{\frac{1}{\xi}(\alpha_1+\alpha_2)}
{\rm sh}\left(\frac{\pi}{\lambda}(\alpha_1-\alpha_2+\pi i)\right)
{\rm sh}\left(\frac{\pi}{\lambda}(\alpha_2-\alpha_1+\pi i)\right)
\nonumber\\
&\times&
{\rm sh}\left(\frac{1}{\xi}(\alpha_1-\alpha_2+\pi i)\right)
{\rm sh}\left(\frac{1}{\xi}(\alpha_1-\beta_2-\pi i)\right)
{\rm sh}\left(\frac{1}{\xi}(\alpha_2-\beta_2-\pi i)\right)
,\nonumber\\
\end{eqnarray}

~\\
and
\begin{eqnarray}
G_2(\beta_1,\beta_2)_{1,1}
&=&e^{-\frac{1}{\xi}(\beta_1+\beta_2)}
\frac{
S_2(i(\beta_2-\beta_1)+\pi|\pi \xi, \lambda)}{
S_2(i(\beta_2-\beta_1)+\pi \xi-\pi|\pi \xi, \lambda)}
\int_{-\infty}^\infty d\alpha_1
\int_{-\infty}^\infty d\alpha_2 \nonumber\\
&\times&\left\{\prod_{j,k=1}^2 \varphi_2(\beta_j-\alpha_k)\right\}
\varphi_1(\alpha_1-\alpha_2)({\cal J}(\alpha_1-\alpha_2+\pi i)
+{\cal J}(\alpha_1-\alpha_2-\pi i))
\nonumber\\
&\times&
e^{\frac{1}{\xi}(\alpha_1+\alpha_2)}
{\rm sh}\left(\frac{\pi}{\lambda}(\alpha_1-\alpha_2+\pi i)\right)
{\rm sh}\left(\frac{\pi}{\lambda}(\alpha_2-\alpha_1+\pi i)\right)
\nonumber\\
&\times&
{\rm sh}\left(\frac{1}{\xi}(\alpha_1-\alpha_2+\pi i)\right)
{\rm sh}\left(\frac{1}{\xi}(\alpha_1-\beta_2-\pi i)\right)
{\rm sh}\left(\frac{1}{\xi}(\alpha_2-\beta_1+\pi i)\right)
.\nonumber
\\
\end{eqnarray}
Here we omit an irrelevant constant factor.\\
By applying Wick's theorem we obtain the $N$-point correlation
functions.
We consider the special components :
\begin{eqnarray}
j_1,\cdots,j_L=2,~
j_{L+1},\cdots,j_{L+2M}=1,~
j_{L+2M+1},\cdots, j_{2(L+M)}=0.
\end{eqnarray}
By using the $R$-matrix symmetry (\ref{C1})
we obtain every components from this component.
The $N$-point function is evaluated as following.
\begin{eqnarray}
G_{2(L+M)}(\beta_1
\cdots \beta_L|\beta_{L+1},\cdots,\beta_{L+2M}|
\beta_{L+2M+1}\cdots \beta_{2(L+M)})_{2,\cdots ,2,
1, \cdots ,1, 0, \cdots ,0}\nonumber\\
=
E(\{\beta\})\int d\alpha ~\Psi(\{\alpha\}|\{\beta\})
~Pf(\{\alpha\})
~I_{\lambda}(\{\alpha\})
~I_\xi(\{\alpha\}|\{\beta\}),
\end{eqnarray}
where the integral $\int d\alpha$
represents
\begin{eqnarray}
\int_{-\infty}^\infty
d\alpha_{L+1} \cdots
\int_{-\infty}^\infty
d\alpha_{2(L+M)}
\int_{-\infty}^\infty
d\alpha_{L+2M+1}' \cdots
\int_{-\infty}^\infty
d\alpha_{2(L+M)}'.
\end{eqnarray}
Here we have set
\begin{eqnarray}
E(\{\beta\})&=&e^{-\frac{1}{\xi}(\beta_{L+1}+\cdots+\beta_{L+2M})}
e^{-\frac{2}{\xi}(\beta_{L+2M+1}'+\cdots+\beta_{2(L+M)}')}
\nonumber\\
&\times&
\prod_{1\leq j<k \leq 2(L+M)}
\frac{S_2(i(\beta_k-\beta_j)+\pi|\pi \xi,\lambda)}
{S_2(i(\beta_k-\beta_j)+\pi \xi-\pi|\pi \xi,\lambda
)}.
\end{eqnarray}
The integral kernel is given by
\begin{eqnarray}
&&\Psi(\{\alpha\}|\{\beta\})=\prod_{L+1\leq j<k \leq 2(L+M)}
\varphi_1(\alpha_j-\alpha_k)
\prod_{L+2M+1\leq j<k \leq 2(L+M)}
\varphi_1(\alpha_j'-\alpha_k')\nonumber\\
&\times&
\prod_{j=L+1}^{2(L+M)}
\prod_{k=L+2M+1}^{2(L+M)}\varphi_1(\alpha_j-\alpha_k')
\prod_{j=1}^{2(L+M)}
\prod_{k=L+1}^{2(L+M)}\varphi_2(\beta_j-\alpha_k)
\prod_{j=1}^{2(L+M)}
\prod_{k=L+2M+1}^{2(L+M)}\varphi_2(\beta_j-\alpha_k').\nonumber\\
\end{eqnarray}
The $Pf(\{\alpha\})$ represents
a Pfaffian of $2(L+M) \times 2(L+M)$ anti-symmmetry
matrix whose entries are given by
${\cal J}(\alpha-\alpha'+\pi i)
+{\cal J}(\alpha-\alpha'-\pi i)
$. \\
The integrand functions are given by
\begin{eqnarray}
&&I_{\lambda}(\{\alpha\})=
\prod_{j=L+1}^{2(L+M)}
\prod_{k=L+1}^{2(L+M)}
{\rm sh}\left(\frac{\pi}{\lambda}(\alpha_j-\alpha_k-\pi i)\right)
\nonumber\\
&\times&\prod_{j=L+1}^{2(L+M)}\prod_{k=L+2M+1}^{2(L+M)}
{\rm sh}\left(\frac{\pi}{\lambda}(\alpha_j-\alpha_k'-\pi i)\right)
{\rm sh}\left(\frac{\pi}{\lambda}(\alpha_j-\alpha_k'+\pi i)\right)
\end{eqnarray}
and
\begin{eqnarray}
&&I_{\xi}(\{\alpha\}|\{\beta\})=
e^{\frac{1}{\xi}(\alpha_{L+1}+\cdots \alpha_{2(L+M)}+
\alpha_{l+2M+1}'+\cdots +\alpha_{2(L+M)}')}\nonumber\\
&\times&
\prod_{L+1\leq j<k \leq 2(L+M)}
{\rm sh}\left(\frac{1}{\xi}(\alpha_j-\alpha_k+\pi i)\right)
\prod_{j=L+1}^{L+2M}\prod_{k=L+2M+1}^{2(L+M)}
{\rm sh}\left(\frac{1}{\xi}(\alpha_j-\alpha_k'+\pi i)\right)
\nonumber\\
&\times&
\prod_{j=L+2M+1}^{2(L+M)}{\rm sh}
\left(\frac{1}{\xi}(\alpha_j-\alpha_k'-{\rm sgn}(j-k)\pi i)\right)
\\
&\times&
\prod_{j=1}^L
\left\{
\prod_{k=L+1}^{2(L+M)}
{\rm sh}\left(\frac{1}{\xi}(\beta_j-\alpha_k-\pi i)\right)
\prod_{k=L+2M+1}^{2(L+M)}
{\rm sh}\left(\frac{1}{\xi}(\beta_j-\alpha_k'-\pi i)\right)
\right\}\nonumber\\
&\times&
\prod_{j=L+1}^{L+2M}
\left\{
\prod_{k=L+2M+1}^{2(L+M)}
{\rm sh}\left(\frac{1}{\xi}(\beta_j-\alpha_k-\pi i)\right)
\prod_{k=L+2M+1}^{2(L+M)}
{\rm sh}\left(\frac{1}{\xi}(\beta_j-\alpha_k'-\pi i)\right)
\right\}\nonumber\\
&\times&
\prod_{j,k=L+1 \atop{j \neq k}}^{L+2M}
{\rm sh}\left(\frac{1}{\xi}(\beta_j-\alpha_k+
{\rm sgn}(j-k)\pi i)\right)
\nonumber\\
&\times&
\prod_{j,k=L+2M+1 \atop{j \neq k}}^{2(L+M)}
\left\{
{\rm sh}\left(\frac{1}{\xi}(\beta_j-\alpha_k+
{\rm sgn}(j-k)\pi i)\right)
{\rm sh}\left(\frac{1}{\xi}(\beta_j-\alpha_k'+
{\rm sgn}(j-k)\pi i)\right)\right\}\nonumber\\
\end{eqnarray}

Here we omit an irrelevant constant factor.\\

Next we consider the special case $\lambda=2\pi$,
in which the trace function will become
the correlation functions of our original solvable
lattice problem.
Note that the special case $\lambda=2\pi$ the kernel
functions simplify.
\begin{eqnarray}
{\cal J}(\alpha+\pi i)+
{\cal J}(\alpha-\pi i)=\delta(\alpha-\pi i),
\end{eqnarray}
and
\begin{eqnarray}
\frac{{\rm tr}_{{\cal H}^b}\left(e^{-\lambda D^b}
U_j(\beta_1)U_j(\beta_2)\right)}{
{\rm tr}_{{\cal H}^b}\left(e^{-\lambda D^b}\right)
}=Const.
\frac{\displaystyle
{\rm ch}\left(\frac{1}{2}(\beta_1-\beta_2)\right)
}{\displaystyle
{\rm sh}\left(
\frac{1}{\xi}(\beta_1-\beta_2-\pi i)\right)
}.
\end{eqnarray}
We have seen that the kernel function of
trace formulae gets simplified when specialized at $\lambda=2\pi$.
Here we summarize the simplified formulae
for the one-point correlation functions
at $\lambda=2\pi$.
In this simplified formulae the  number of the contour
integrals reduces to one.
This is due to a property of the fermion
two-points function which becomes the delta function.

~\\
\begin{eqnarray}
&&G_{2}(\beta_1,\beta_2)_{2,0}=
e^{-\frac{2}{\xi}\beta_2}
\frac{\displaystyle
{\rm ch}\left(\frac{1}{2}(\beta_1-\beta_2)\right)}
{
\displaystyle
{\rm sh}\left(\frac{1}{\xi}(\beta_1-\beta_2-\pi i)\right)
}\nonumber\\
&\times&
\int_{-\infty}^\infty d\alpha
\frac{e^{\frac{2}{\xi}\alpha}}
{ \prod_{k=1}^2
\displaystyle
{\rm sh}\left(
\beta_k-\alpha\right)
}
{\rm sh}\left(\frac{1}{\xi}
(\beta_1-\alpha-\pi i)\right)
{\rm sh}\left(\frac{1}{\xi}
(\beta_1-\alpha-2\pi i)\right).
\end{eqnarray}

\begin{eqnarray}
&&G_{2}(\beta_1,\beta_2)_{0,2}=
e^{-\frac{2}{\xi}\beta_1}
\frac{\displaystyle
{\rm ch}\left(\frac{1}{2}(\beta_1-\beta_2)\right)}
{
\displaystyle
{\rm sh}\left(\frac{1}{\xi}(\beta_1-\beta_2-\pi i)\right)
}\nonumber\\
&\times&
\int_{-\infty}^\infty d\alpha
\frac{e^{\frac{2}{\xi}\alpha}}
{ \prod_{k=1}^2
\displaystyle
{\rm sh}\left(
\beta_k-\alpha\right)
}
{\rm sh}\left(\frac{1}{\xi}
(\beta_2-\alpha)\right)
{\rm sh}\left(\frac{1}{\xi}
(\beta_2-\alpha+ \pi i)\right).
\end{eqnarray}

\begin{eqnarray}
&&G_{2}(\beta_1,\beta_2)_{1,1}=
e^{-\frac{1}{\xi}(\beta_1+\beta_2)}
\frac{\displaystyle
{\rm ch}\left(\frac{1}{2}(\beta_1-\beta_2)\right)}
{
\displaystyle
{\rm sh}\left(\frac{1}{\xi}(\beta_1-\beta_2-\pi i)\right)
}\nonumber\\
&\times&
\int_{-\infty}^\infty d\alpha
\frac{e^{\frac{2}{\xi}\alpha}}
{ \prod_{k=1}^2
\displaystyle
{\rm sh}\left(
\beta_k-\alpha\right)
}
{\rm sh}\left(\frac{1}{\xi}
(\beta_1-\alpha-\pi i)\right)
{\rm sh}\left(\frac{1}{\xi}
(\beta_2-\alpha)\right).\\
\nonumber
\end{eqnarray}
Here we omit an irrelevant constant factor.
The $R$-matrix symmetry (\ref{C1}) of the 
above integral representations 
can be reduced to the special case of the identity (\ref{weak}):
$\alpha_1=\alpha+\pi i,~\alpha_2=\alpha$.
The cyclicity condition (\ref{C2})
can be checked by straightforward calculation.
We have seen that the formulae of the one-point functions
get simplified when we set $\lambda=2\pi$.
This feature holds for the $N$-point correlation functions,
too.
The number of the contour integrals reduces to only $N$.

~\\
~\\
{\bf Acknowledgements.}~~
This work was partly supported by Grant-in-Aid for
Encouragements for Young Scientists ({\bf A})
from Japan Society for the Promotion of Science. (11740099)

\begin{appendix}
\section{Multi Gamma functions}
Here we summarize the multiple gamma and the multiple sine
functions, following N.Kurokawa \cite{Kur}.\\
Let us set the functions
$\Gamma_1(x|\omega)$ and $\Gamma_2(x|\omega_1, \omega_2)$
by
\begin{eqnarray}
{\rm log}\Gamma_1(x|\omega)+\gamma B_{11}(x|\omega)&=&
\int_C\frac{dt}{2\pi i t}e^{-xt}
\frac{{\rm log}(-t)}{1-e^{-\omega t}},\\
{\rm log}\Gamma_2(x|\omega_1, \omega_2)
-\frac{\gamma}{2} B_{22}(x|\omega_1, \omega_2)&=&
\int_C\frac{dt}{2\pi i t}e^{-xt}
\frac{{\rm log}(-t)}{(1-e^{-\omega_1 t})
(1-e^{-\omega_2 t})},
\end{eqnarray}
where
the functions $B_{jj}(x)$ are the multiple Bernoulli polynomials
defined by
\begin{eqnarray}
\frac{t^r e^{xt}}{
\prod_{j=1}^r (e^{\omega_j t}-1)}=
\sum_{n=0}^\infty
\frac{t^n}{n!}B_{r,n}(x|\omega_1 \cdots \omega_r),
\end{eqnarray}
more explicitly
\begin{eqnarray}
B_{11}(x|\omega)&=&\frac{x}{\omega}-\frac{1}{2},\\
B_{22}(x|\omega)&=&\frac{x^2}{\omega_1 \omega_2}
-\left(\frac{1}{\omega_1}+\frac{1}{\omega_2}\right)x
+\frac{1}{2}+\frac{1}{6}\left(\frac{\omega_1}{\omega_2}
+\frac{\omega_2}{\omega_1}\right).
\end{eqnarray}
Here $\gamma$ is Euler's constant,
$\gamma=\lim_{n\to \infty}
(1+\frac{1}{2}+\frac{1}{3}+\cdots+\frac{1}{n}-{\rm log}n)$.\\
Here the contor of integral is given by

~\\
~\\

\unitlength 0.1in
\begin{picture}(34.10,11.35)(17.90,-19.35)
%
\special{pn 8}%
\special{pa 5200 800}%
\special{pa 2190 800}%
\special{fp}%
\special{sh 1}%
\special{pa 2190 800}%
\special{pa 2257 820}%
\special{pa 2243 800}%
\special{pa 2257 780}%
\special{pa 2190 800}%
\special{fp}%
\special{pa 2190 1600}%
\special{pa 5190 1600}%
\special{fp}%
\special{sh 1}%
\special{pa 5190 1600}%
\special{pa 5123 1580}%
\special{pa 5137 1600}%
\special{pa 5123 1620}%
\special{pa 5190 1600}%
\special{fp}%
%
\special{pn 8}%
\special{pa 5190 1200}%
\special{pa 2590 1210}%
\special{fp}%
\put(25.9000,-12.1000){\makebox(0,0)[lb]{$0$}}%
%
\special{pn 8}%
\special{ar 2190 1210 400 400  1.5707963 4.7123890}%
\put(33.9000,-20.2000){\makebox(0,0){{\bf Contour} $C$}}%
\end{picture}%

~\\
~\\

Let us set
\begin{eqnarray}
S_1(x|\omega)&=&\frac{1}{\Gamma_1(\omega-x|\omega)
\Gamma_1(x|\omega)},\\
S_2(x|\omega_1,\omega_2)&=&\frac{
\Gamma_2(\omega_1+\omega_2-x|\omega_1,\omega_2)}{
\Gamma_2(x|\omega_1,\omega_2)}.
\end{eqnarray}
We have
\begin{eqnarray}
\Gamma_1(x|\omega)=e^{(\frac{x}{\omega}-\frac{1}{2}){\rm log}
\omega}\frac{\Gamma(x/\omega)}{\sqrt{2\pi}},~
S_1(x|\omega)=2{\rm sin}(\pi x/\omega),
\end{eqnarray}
\begin{eqnarray}
\frac{\Gamma_2(x+\omega_1|\omega_1,\omega_2)}{
\Gamma_2(x|\omega_1,\omega_2)}=\frac{1}{\Gamma_1(x|\omega_2)},~
\frac{S_2(x+\omega_1|\omega_1,\omega_2)}{
S_2(x|\omega_1,\omega_2)}=\frac{1}{S_1(x|\omega_2)},~
\frac{\Gamma_1(x+\omega|\omega)}{\Gamma_1(x|\omega)}=x.
\end{eqnarray}

\begin{eqnarray}
{\rm log}S_2(x|\omega_1 \omega_2)
=\int_C \frac{{\rm sh}(x-\frac{\omega_1+\omega_2}{2})t}
{2{\rm sh}\frac{\omega_1 t}{2}
{\rm sh}\frac{\omega_2 t}{2}
}{\rm log}(-t)\frac{dt}{2\pi i t},~(0<{\rm Re}x<
\omega_1+\omega_2).
\end{eqnarray}

\begin{eqnarray}
S_2(x|\omega_1 \omega_2)=
\frac{2\pi}{\sqrt{\omega_1 \omega_2}}x +O(x^2),~~(x \to 0).
\end{eqnarray}

\begin{eqnarray}
S_2(x|\omega_1 \omega_2)
S_2(-x|\omega_1 \omega_2)=-4
{\rm sin}\frac{\pi x}{\omega_1}
{\rm sin}\frac{\pi x}{\omega_2}.
\end{eqnarray}

\end{appendix}

\end{document}